\documentstyle[12pt,draft,epsf]{article}
\def\eq{\begin{equation}}
\def\en{\end{equation}}
\newcommand \be  {\begin{equation}}
\newcommand \bea {\begin{eqnarray} \nonumber }
\newcommand \ee  {\end{equation}}
\newcommand \eea {\end{eqnarray}}

\def\B{\beta}

\def \bi{\bibitem}

\def\d{{\rm d}}
 \def\(({\left(}
 \def\)){\right)}
\def\bi{\bibitem}

\def \ov{\over}

\def \a{\alpha}
\def \b{\beta}
\def\D{\Delta}
\def \z{\zeta}

\def \d{{\rm d}}
\def \e{{\rm e}}
\def \del{\delta}

\def \beqna{\begin{eqnarray}}
\def \eeqna{\end{eqnarray}}
\def \beq{\begin{equation}}
\def \eeq{\end{equation}}
\def \be{\begin{equation}}
\def \ee{\end{equation}}

\def \ov{\over}

\def \ol{\overline}
\def \a{\alpha}

\def \b{\beta}

\def \la{\langle}
\def \ra{\rangle}

\def \ab2{\alpha\beta^2}

\def \la{\langle}
\def \ra{\rangle}

\begin{document}

\baselineskip25pt

%\begin{center}

%{\LARGE\bf  }
%\vskip1cm

%\end{center}

\baselineskip20pt

\begin{center}
{\Large {\bf Dynamical solution of a model without energy barriers}}\\
{\Large Silvio Franz (1), Felix Ritort (2)}
\vskip1cm

{\large
(1)  NORDITA and CONNECT\\
     Blegdamsvej 17,\\
     DK-2100 Copenhagen \O\\
     Denmark\\
e-mail: {\it franz@nordita.dk }\\ }

\vskip0.5cm
{\large
(2) Departamento de Matematicas,\\
Universidad Carlos III, Butarque 15\\
Legan\'es 28911, Madrid (Spain)\\
e-mail: {\it ritort@dulcinea.uc3m.es}\\}
\vskip0.5cm

{\large May 1995}
\vskip1cm

\end{center}

\baselineskip16pt

\vskip1cm

{\bf Abstract:} In this note we study the dynamics of a model recently
introduced by one of us, that displays glassy phenomena in absence of
energy barriers. Using an adiabatic hypothesis we derive an equation for
the evolution of the energy as a function of time that describes
extremely well the glassy behaviour
observed in   Monte Carlo simulations.

\vskip1cm

\begin{flushright}
cond-mat/9505115
\end{flushright}

\vfill
\eject

Despite intensive studies both on the experimental and the theoretical
sides, the mechanisms responsible for the appearance of the glassy state
are far from being understood.  Experiments show that the relaxation
time of supercooled liquids increases dramatically as the temperature is
lowered, and the glass is formed when it exceeds the probing time
\cite{G}.  This slowing down is related to the appearance of high
free-energy barriers.  Examples of simple models with a glassy
dynamics resulting from high free-energy barriers have been studied in
detail and can be found in \cite{BoMe,CuKu,MaPaRi}.  Free-energy
barriers can be thought as composed of an energetic contribution (the
landscape defined by the Hamiltonian), and an entropic one related to
the small number of paths that from a given state lead to the thermal
equilibrium.  In order to understand the importance of the entropy
barriers in the formation of glasses, it has been recently considered a
model which has not energy barriers, but that presents glassy phenomena
due to entropy barriers \cite{I}.

In this note we study analytically the relaxation dynamics of the
model \cite{I} (hereafter referred as I).

In the model there are $N$ distinguishable particles that can occupy $N$
different states $r=1,...,N$.  The energy of a configuration is given by
the number of unoccupied sites, or defining the occupation numbers
of the states $n_r$,
\be
H=-\sum_{r=1}^N \del_{n_r,0}.
\ee
We consider the sequential Metropolis dynamics such that at each
Monte Carlo sweep a particle is chosen at random and it is moved to a
random arrival state with probability one if the energy does not
increase, and with probability $exp(-\B)$ if the energy increases of a
unity.  It is worth noticing that the only processes decreasing the
energy are the ones in which particles coming from departure states with
$n_d=1$ fall in arrival states with $n_a>0$. The energy variation of a
single sweep can be zero or $\pm 1$. Note that in the model there are
not energy barriers: the ground state can be always reached decreasing
the energy monotonically. The ground state corresponds to all particles
occupying one state, hence its energy per particle (or state) is $-1$.
The dynamics of this model is very slow at low temperatures because, as
the energy decreases with time, it takes always a larger time to the
system to empty one more state. Due to the particular rules of the
dynamics it has been suggested to call it backgammon  (BG) model and we
will adopt this name hereafter\footnote{We are grateful to
Marc Potters for this suggestion.}.

In order to understand the relaxation dynamics of the BG model, it is
useful to study the canonical probability distribution of the occupation
numbers at equilibrium. An elementary calculation, along the lines of I,
shows that for large $N$ the probability of the different occupation
numbers factorizes, and the single number distribution is given by the
following modified Poisson Law:
\be
P_\b(n)=\frac{1}{\e^{\zeta}\zeta} \e^{\b \del_{n,0}}{\zeta^n\ov n!}
\label{distr}
\ee
where the `fugacity' $\zeta$ is determined self-consistently by the
solution of the equation:
\be
{\zeta \e^\zeta\ov \e^\b +\e^\z-1}=1.
\label{rel}
\ee
The internal energy is easily obtained as $U=-P_\b(n=0)$.

The basic observation that allows us to solve the long-time dynamics of
the BG model is that at low temperatures the distribution (\ref{distr}) has
a gap at low ($>0$) occupation numbers.  Therefore it is conceivable
that in the low-temperature dynamics the degrees of freedom that lead to
the Poisson distribution equilibrate much faster than the energy
itself. During the dynamical evolution, the moves that change
the energy are rare respect to the moves that lead to the equilibration
of the probability distribution of the occupation numbers
at fixed energy.
Hence, at low temperatures, we expect the probability distribution of the
occupation numbers to follow (at large enough times) a law of the kind
eq.(\ref{distr}) to a high degree of approximation, with $\b$ and
$\zeta$ related by (\ref{rel}) as effective time-dependent parameters.

This observation allows us to close the dynamical equation for the
evolution of the energy $U(t)$, and consequently also for the occupation
numbers probability distribution itself. It will result a theory able
to predict the evolution of all the  quantities depending just on that
probability distribution.

An adiabatic hypothesis similar to the one used here
was used as an approximation by Coolen and
Sherrington to study the relaxation of the energy (and some other quantities)
in spin glass and neural network models  \cite{CS}. The present
work puts that approximation
in perspective: we believe it to be valid when
only entropy, and not energy barriers, are the responsible for the slowing
down of the dynamics.

Let us consider a single Monte Carlo sweep, corresponding
to an infinitesimal time interval $\del t=1/N$; this is  specified by
$3$ independent random variables:
\begin{itemize}
\item{A state of departure $d=1,...,N$, to be chosen
with probability $n_d/N$.\footnote{This corresponds to the uniform
choice of a particle to move.}}
\item{An arrival state $a=1,...,N$ with a uniform probability $\frac{1}{N}$}
\item{An acceptance variable $x$ equal to 1 with probability
$\e^{-\b}$ and zero otherwise, determining the acceptance of the move
if it increases the energy.}
\end{itemize}
In terms of this variables the elementary energy variation can be written as
\be
\D E(t)= -\del_{n_d,1} (1-\del_{n_a,0})+(1-\del_{n_d,1})\del_{n_a,0}\del_{x,1}
\ee
Observing that $1/N \sum_a \del_{n_a,0} =-U(t)$ we find that
the average over the distributions of these variables reads
\be
[U(t+\del t)-U(t)]N\equiv {\d U\ov \d t}=
-\Bigl(1+U(t)\Bigr)P(n=1,t)-\e^{-\b}U(t)(1-P(n=1,t)).
\label{exact}
\ee
The previous expression is an exact relation. A closed equation
can be obtained substituting the exact probability distribution
$P(n=1,t)$ at the time $t$ with the modified Poisson
distribution (\ref{distr}) for $P_{\b^*(t)}(n=1)$
with parameters $\z^*(t)$ and
$\b^*(t)$ determined self-consistently  by the relations
\be
U(t)=-P_{\b^*(t)}(n=0)\;\;\;\;\;;\;\;\;
{ \zeta^*(t) \e^{\zeta^*(t)}\ov \e^{\b^*(t)} +\e^{\z^*(t)}-1 }=1.
\label{reldyn}
\ee

The previous equation gives the expected result ${\d U\ov \d t}=0$ at
equilibrium ($\zeta^*(t)=\zeta,\b^*(t)=\b$). We note that the relaxation
eq.(\ref{exact}) is of the Markovian type; memory effects
are not important in the relaxation of the energy. The basic mechanism
of equilibration of the occupation number distribution on the constant
energy surface is not expected to hold  at very early stages of the dynamics
(where the energy changes very fast). Coherently we find discrepancies
with Monte Carlo simulations at short times, but excellent agreement
for large times. The linearization of eq. (\ref{exact},\ref{reldyn})
around the fixed point $U=U_{eq}$ ($\b^*=\b$) allows to study the
exponential relaxation time for the energy.
In terms of $\z$ as a solution of eq.(\ref{rel}),
the relaxation time reads:
$\tau=  \big[1+(\z-1)\e^\z\big]\big[1+\z -
\e^\z\big]/\big[\z^2(1-\e^z)\big]$.
For large $\b$ it is found $\tau\sim   \e^\b/\b^2$.

The relaxation equation (\ref{exact}) is particularly simple at zero
temperature. In this case, only processes which
decrease the energy contribute to the rate of variation of the energy. It
 is easy to observe that $\frac{dU}{dt}$ goes like
$exp(-\beta^*(t))$. Because $U(T^*)=-1+T^*$ at small temperatures $T^*$
this yields,
\be
\frac{dU}{dt}=exp\Bigl(\frac{-1}{1+U}\bigr)
\ee

in agreement with the conjectured expression in I.
 The results for the decay of the energy
at zero temperature are shown in figure 1.
\begin{figure}
% \epsffile[120 206 565 440]{eneT0epsf1.ps}
%\epsffile[0 206 565 440]{eneT0epsf1.ps}
\epsffile[0 0 565 440]{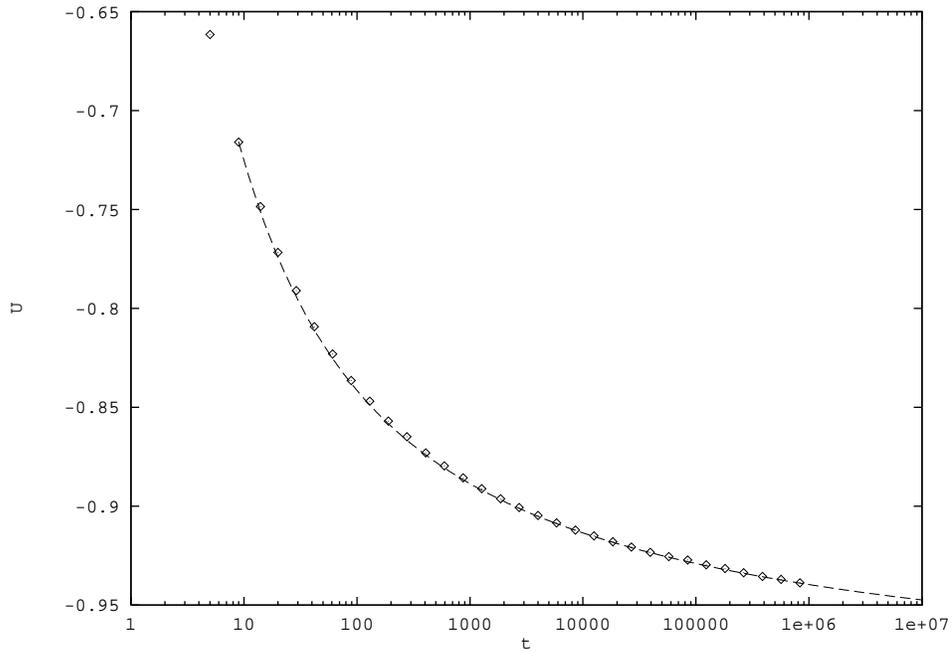}
\caption[a]{{ \protect
%\small
 Energy as a function of time compared to Monte Carlo
data for $N=20000$ at temperatures $T=0$. We do not show the results
of (\ref{exact},\ref{reldyn}) for very short times, where appreciable
differences with the Monte Carlo data are observed
}}
\end{figure}
 These have been obtained from
the the numerical integration of equation (\ref{exact}) and from the
Monte Carlo dynamics. The agreement is very good especially for large
times.

Furthermore, equation ($\ref{exact}$) is simple enough to allow for the
study of the dependence of the energy on the cooling (and also heating)
rate. This is shown in figure 2 where we plot the numerical integration
of equation $(\ref{exact})$ with the Monte Carlo data for different
cooling rates (the cooling rate $\a$ is defined as the variation of
temperature per unit time in the dynamical process).
\begin{figure}
\epsffile[0 0 565 440]{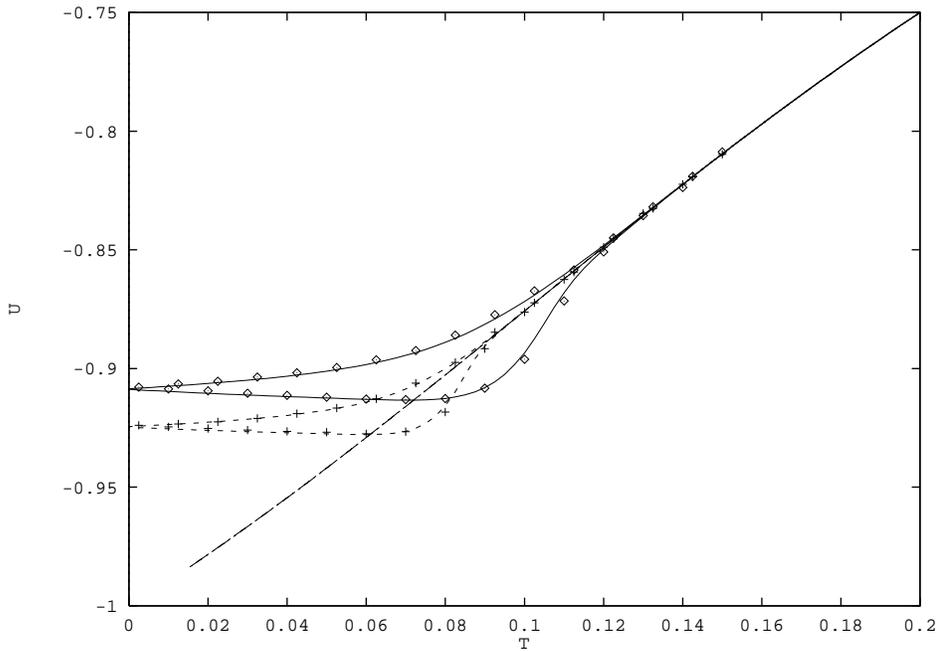}
\caption{{
%\small
Hysteresis cycles obtained  integrating numerically
equation (5,6)
%(\ref{exact})
 for different cooling-heating rates.
The cooling-heating rates are $\a=3.3\times 10^{-5} MCstep^{-1}$ (lines with
diamonds) and
 $\a=3.3\times 10^{-6} MCstep^{-1}$ (lines with crosses). The points are the
Monte Carlo data for $N=20000$.
}}
\end{figure}
 Similar results
have been obtained for the dependence of the energy on the rate
variation of the temperature during a heating process starting from the
ground state configuration. All these results are in
agreement with those presented in I and prove that hysteresis effects
are very strong in this model. The appearance of hysteresis loops is
shown in figure 2 where the energy is plotted as a function of the
temperature during a cooling-heating process. Similar to what has been
observed in experiments on real glasses, the enclosed area of the
hysteresis loop decreases with the cooling-heating rates. One striking
feature in figure 2 is that the dynamical energy in the heating process
goes below the equilibrium one. This is a general feature observed in real
glasses \cite{Se}. As far as we know this has never
been observed in spin glasses
\cite{MaPaRi}.

We tried to interpret the decay of the energy autocorrelation function
defined in I along the same lines,
\be
C_U(t,s)=\frac{\frac{1}{N}\sum_r\delta_{n_r(t),0}\delta_{n_r(s),0}
-U(t)U(s)}{-U(s)(1+U(s))}\,\,\,\,t\ge s
\label{correla}
\ee

This function displays aging at low temperatures and has been used in I
in order to estimate the relaxation time.  Unfortunately, finding the
solution for $C_U(t,s)$ proves to be a much difficult task as not only
the processes of emptying the states are relevant, but also the
diffusion of `towers' of particles from site to site contribute.
Neglection of the diffusion mechanism would lead to the crude estimate
$C_U(t,s)\sim [1+U(t)]/[1+U(s)]$ which  does not fit the
simulation data.  In order to disentangle the two effects we considered
the micro-canonical dynamics, where only moves that do not variate the
energy $U$ are taken into account. In this case, the occupation numbers
distribution thermalizes very fast while the diffusion becomes uniform
in time. The correlation function $C_U(t,s)$ (which coincides with the
canonical equilibrium one at the temperature $T_U$ through the equilibrium
relation (\ref{rel})) is time-translation invariant.  Monte Carlo
simulations in the microcanonical ensemble show that $C_U(t,s)$ is
 a simple exponential. We conclude that the aging in the
function (\ref{correla}) is due to the interplay of the two
aforementioned effects.

 Let
us consider the evolution of the quantity
\be
M_0(t,s)=\frac{1}{N}\sum_r\delta_{n_r(t),0}\delta_{n_r(s),0}~.
\ee
Contributions to $M_0(t,s)$ come from both the emptying of the states
and the diffusion of the towers. In the same spirit as it
has been done for
the energy, we can write the stochastic equation for the
variation of $M_0(t,s)$ in an elementary move of a particle. We
consider the zero temperature case (i.e. $x=0$) and (using the same
notation as for the energy) we get,

\be
N(\Delta_t M_0(t,s))=\delta_{n_d(t),1}(-\delta_{n_a(t),0}\delta_{n_a(s),0}
+\delta_{n_d(s),0})~.
\label{eqT0}
\ee

Averaging over the distributions of the variables $n_a$ and $n_d$ we
obtain
\be
\frac{dM_0(t,s)}{dt}=-P(n=1,t)M_0(t,s)+M_1(t,s)
\label{eqdT0}
\ee

where $M_1(t,s)=\langle\delta_{n_r(t),1}\delta_{n_r(s),0}\rangle$ and
$P(n=1,t)$ is the average number of states with one particle at time
$t$. It is clear that considering the correlations
$M_k(t,s)=\langle\delta_{n_r(t),k}\delta_{n_r(s),0}\rangle$,  it is
possible to obtain in a recursive way a hierarchical set of first-order
differential equations specifying $M_k(t,s)$ as a function of
$M_{k+1}(t,s)$. Approximations can be obtained truncating the hierarchy
at some order $k$ and writing $M_{k+1}(t,s)\simeq
-P(n=k,t)U(s)$. Although we did not try a systematic study, we expect
that good approximations could be obtained truncating the hierarchy at a
value of $k$ large enough that $P(n=k,t)$ is negligible for large
$t$. Unfortunately, truncation at $k=1$ gives very poor results for the
correlation function.

Equations (\ref{exact},\ref{reldyn}) enable us to study the response
function of the system defined as $r_U(t,s)={\del U(t)\ov \del \b(s)}$
conjugated at equilibrium to the correlation function
$\ol{C_U}(t,s)=(1/N^2)\sum_{r,r'}\la
\del_{n_r(t),0}\del_{n_{r'}(s),0}\ra$ via the fluctuation-dissipation
theorem.\footnote{Note the difference of definition among $\ol{C}$ and
$M_0$, in the last quantity terms of the type
$\langle\delta_{n_r(t),k}\delta_{n_r'(s),0}\rangle$ with $r\ne r'$ are
absent.}  In order to study $r_U$ we have emulated an aging experiment
(solving equation (\ref{exact})). After a quench from high temperature
at time zero the system thermalizes during a waiting time $t_w$ (with
$\b$ kept fixed), then the temperature is changed to $\b+\del \b$,
choosing $\del \b$ small enough to be in the linear response regime.
We then computed the difference of the energies at the time $t>t_w$
obtained with this last procedure and the one at $\b$ fixed

\be
\chi_U(t,t_w)={U(t,\b+\del\b)-U(t,\b)\ov \del \b} =\int_{t_w}^t \d
s\,r_U(t,s).
\ee

In figure 3 we plot this quantity for different waiting times and
temperatures as a function of $t-t_w$. It is apparent that
only a very weak dependence on the
waiting time is present, i.e. $\chi_U(t,t_w)\simeq\chi_U(t-t_w)$.  The
large-time limit of $\chi_U(t,t_w)$ is the one predicted by the
fluctuation-dissipation theorem $\chi_U^{eq}(t)=-T^2 C_v\equiv
\partial U/\partial \b$ where $C_v$ is the equilibrium specific heat. It
would be interesting to compute the correlation function $\ol{C_U}$ to
investigate if the fluctuation-dissipation theorem is verified at all
times. Unfortunately eqs.(\ref{exact},\ref{reldyn}) do not give access to
the response function conjugated to $M_0(t,s)$ which is the correlation
that exhibits aging.
\begin{figure}
% \epsffile[120 206 565 440]{eneT0epsf1.ps}
%\epsffile[0 206 565 440]{eneT0epsf1.ps}
\epsffile[0 0 565 440]{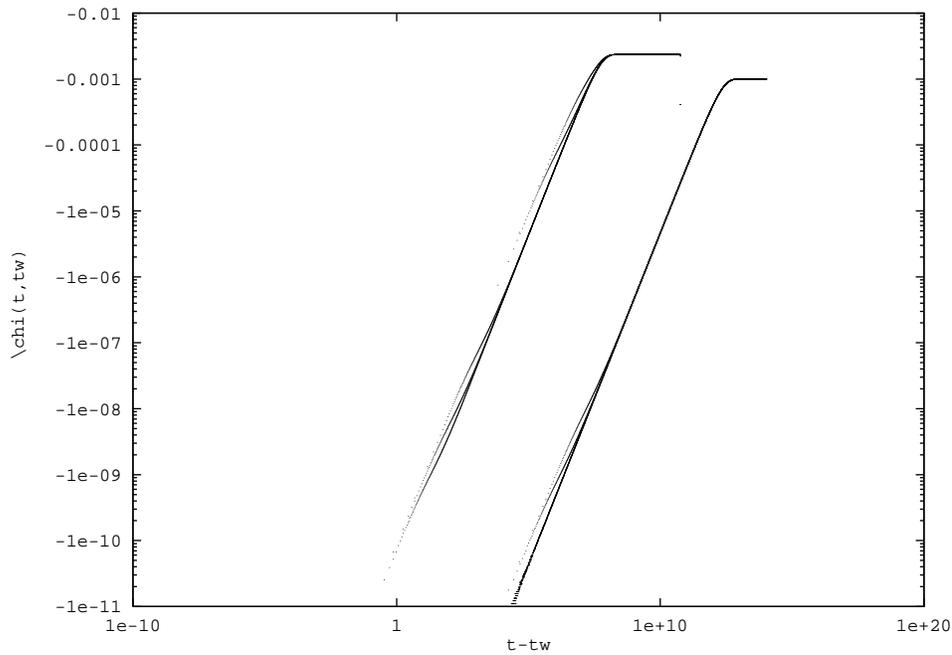}
\caption[a]{{ \protect
%\small
Integrated response function for different temperatures
$T=2.9\times 10^{-2}$ (right curve),
$T=4.3\times 10^{-2}$ (left curve)  and
different waiting times (ranging from $t_w=20$ to $t_w=10^7$,
as a function of $t-t_w$. We see that for
different temperatures the response function behaves roughly
as a power law $\chi(t-t_w)\simeq -a(\b)(t-t_w)^\nu$ in  a large
range of time scales before saturating to its limiting value $T^2 C_v$.
The exponent $\nu$ takes the value $\nu\simeq 1$ nearly independent
of the temperature.
}}
\end{figure}

Summarizing, we have seen that some glassy features (slow relaxation and
hysteresis effects) of the dynamics of the BG model can be understood
under the simple hypothesis of partial equilibrium of the relevant
degrees of freedom of the system on the surface of constant energy.
A more refined theory would be necessary to understand the aging effect
in the correlation function (\ref{correla}) and possibly in its
conjugated response function.

We expect the adiabatic mechanism described in this paper to be relevant
whenever the entropic barriers are responsible for the slowing down of
the dynamics.  In this line, it would be interesting to study other
 models where this result holds and variants of the BG model
like for instance, the Hamiltonians
$H=-\frac{1}{N^{p-1}}(\sum_r\,\delta_{n_r,0})^p$ which present a static
phase transition \cite{FraRi}. In those cases, a similar theoretical
analysis in the low $T$ region to that presented in case of the BG model
should be valid.

\vspace{1 cm}
{\bf Acknowledgements}
F.R. acknowledges Nordita for its kind hospitality during the realization
of this work and Universidad Carlos III de Madrid for financial support.

\end{document}